\documentclass[aps,prl,reprint,superscriptaddress,amsmath,amssymb]{revtex4-2}

\usepackage{graphicx}
\usepackage{color}

\newcommand{\etap}{\eta^{\prime}}

\RequirePackage{lineno}

\begin{document}


\title{Direct measurement of the in-medium $\eta^{\prime}$ mass spectrum through the $\gamma\gamma$ decay channel}


\newcommand{\RARiS}{Research Center for Accelerator and Radioisotope Science, Tohoku University, Sendai, Miyagi 982-0826, Japan}
\newcommand{\CAS}{Institute of Modern Physics, Chinese Academy of Sciences, Lanzhou 730000, China}
\newcommand{\RCNP}{Research Center for Nuclear Physics, Osaka University, Ibaraki, Osaka 567-0047, Japan}
\newcommand{\KRU}{Department of Physics, Korea University, Seoul 02841, Republic of Korea}
\newcommand{\SINICA}{Institute of Physics, Academia Sinica, Taipei 11529, Taiwan}
\newcommand{\NSRRC}{National Synchrotron Radiation Research Center, Hsinchu 30076, Taiwan}
\newcommand{\KyotoU}{Department of Physics, Kyoto University, Kyoto 606-8502, Japan}
\newcommand{\NJU}{Department of Nuclear Science and Technology, College of Materials Science and Technology,\\Nanjing University of Aeronautics and Astronautics, Nanjing 210016, China}
\newcommand{\OHU}{Department of Physics and Astronomy, Ohio University, Athens, OH 45701, USA}
\newcommand{\RIKEN}{RIKEN SPring-8 Center, Sayo, Hyogo 679-5148, Japan}
\newcommand{\JLab}{Thomas Jefferson National Accelerator Facility, Newport News, Virginia 23606, USA}
\newcommand{\TokyoU}{Department of Physics, University of Tokyo, Tokyo 113-0033, Japan}
\newcommand{\GifuU}{Department of Education, Gifu University, Gifu 501-1193, Japan}
\newcommand{\DNRI}{Dalat Nuclear Research Institute, Dalat, Lam Dong, Vietnam}
\newcommand{\KSU}{Department of Physics, Kyoto Sangyo University, Kyoto 603-8555, Japan}
\newcommand{\TokyoHU}{Department of Radiology, The University of Tokyo Hospital, Tokyo 113-8655, Japan}
\newcommand{\KEKN}{Institute of Particle and Nuclear Studies,\\High Energy Accelerator Research Organization (KEK), Tsukuba, Ibaraki 305-0801, Japan}
\newcommand{\SKU}{Department of Physics and Engineering Physics, University of Saskatchewan, Saskatoon, Canada SK S7N 5E2}
\newcommand{\JINR}{Laboratory of High Energy Physics, Joint Institute for Nuclear Research, Dubna, Moscow Region 142281, Russia}
\newcommand{\SPr}{Japan Synchrotron Radiation Research Institute (SPring-8), Sayo, Hyogo 679-5198, Japan}
\newcommand{\JAEA}{J-PARC Center, Japan Atomic Energy Agency, Tokai, Ibaraki 319-1195, Japan}
\newcommand{\BHU}{School of Physics, Beihang University, Beijing 100191, China}
\newcommand{\KEKR}{Radiation Science Center, High Energy Accelerator Research Organization (KEK), Tokai, Ibaraki 319-1195, Japan}
\newcommand{\KEKT}{Instrumentation Technology Development Center,\\High Energy Accelerator Research Organization (KEK), Tsukuba, Ibaraki 305-0801, Japan}

\author{Y.~Matsumura}
\affiliation{\RARiS}
\author{N.~Muramatsu}
\affiliation{\CAS}
\author{T.A.~Hashimoto}
\affiliation{\RCNP}
\author{M.~Miyabe}
\affiliation{\RARiS}
\author{H.~Shimizu}
\affiliation{\RARiS}
\author{A.O.~Tokiyasu}
\affiliation{\RARiS}
\author{J.K.~Ahn}
\affiliation{\KRU}
\author{W.C.~Chang}
\affiliation{\SINICA}
\author{J.Y.~Chen}
\affiliation{\NSRRC}
\author{M.L.~Chu}
\affiliation{\SINICA}
\author{S.~Dat\'{e}}
\affiliation{\RCNP}
\author{T.~Gogami}
\affiliation{\KyotoU}
\author{H.~Hamano}
\affiliation{\RCNP}
\author{Q.H.~He}
\affiliation{\NJU}
\author{K.~Hicks}
\affiliation{\OHU}
\author{T.~Hiraiwa}
\affiliation{\RIKEN}
\author{Y.~Honda}
\affiliation{\RARiS}
\author{T.~Hotta}
\affiliation{\RCNP}
\author{Y.~Inoue}
\affiliation{\RARiS}
\author{T.~Ishikawa}
\affiliation{\RCNP}
\author{I.~Jaegle}
\affiliation{\JLab}
\author{Y.~Kasamatsu}
\affiliation{\RCNP}
\author{H.~Katsuragawa}
\affiliation{\RCNP}
\author{S.~Kido}
\affiliation{\RARiS}
\author{R.~Kobayakawa}
\affiliation{\RCNP}
\author{Y.~Kon}
\affiliation{\RCNP}
\author{S.~Masumoto}
\affiliation{\TokyoU}
\author{K.~Mizutani}
\affiliation{\RCNP}
\author{T.Z.~Mo}
\affiliation{\NJU}
\author{T.~Nakamura}
\affiliation{\GifuU}
\author{T.~Nakano}
\affiliation{\RCNP}
\author{T.~Nam}
\affiliation{\DNRI}
\author{M.~Niiyama}
\affiliation{\KSU}
\author{Y.~Nozawa}
\affiliation{\TokyoHU}
\author{Y.~Ohashi}
\affiliation{\RCNP}
\author{H.~Ohnishi}
\affiliation{\RARiS}
\author{T.~Ohta}
\affiliation{\TokyoHU}
\author{M.~Okabe}
\affiliation{\RARiS}
\author{K.~Ozawa}
\affiliation{\KEKN}
\author{C.~Rangacharyulu}
\affiliation{\SKU}
\author{S.Y.~Ryu}
\affiliation{\RCNP}
\author{Y.~Sada}
\affiliation{\RARiS}
\author{T.~Shibukawa}
\affiliation{\TokyoU}
\author{R.~Shirai}
\affiliation{\RARiS}
\author{K.~Shiraishi}
\affiliation{\RARiS}
\author{E.A.~Strokovsky}
\affiliation{\JINR}
\author{Y.~Sugaya}
\affiliation{\RCNP}
\author{M.~Sumihama}
\affiliation{\GifuU}
\affiliation{\RCNP}
\author{S.~Suzuki}
\affiliation{\SPr}
\affiliation{\RCNP}
\author{S.~Tanaka}
\affiliation{\RCNP}
\author{Y.~Taniguchi}
\affiliation{\RARiS}
\author{N.~Tomida}
\affiliation{\KyotoU}
\author{Y.~Tsuchikawa}
\affiliation{\JAEA}
\author{T.~Ueda}
\affiliation{\RARiS}
\author{T.F.~Wang}
\affiliation{\BHU}
\author{H.~Yamazaki}
\affiliation{\KEKR}
\author{R.~Yamazaki}
\affiliation{\RARiS}
\author{Y.~Yanai}
\affiliation{\RCNP}
\author{T.~Yorita}
\affiliation{\RCNP}
\author{C.~Yoshida}
\affiliation{\KEKT}
\author{M.~Yosoi}
\affiliation{\RCNP}
\collaboration{LEPS2/BGOegg collaboration}



\date{\today}

\begin{abstract}
We measured the invariant mass of the $\eta^{\prime}(958)$ meson through the $\eta^{\prime}\to\gamma\gamma$ decay channel to search for possible character change of $\eta^{\prime}$ in a nuclear medium. The measured invariant-mass spectra were analyzed by fitting realistic spectral functions. 
An enhancement was observed in the lower tail of the $\eta^{\prime}$ mass peak for low-momentum ($P_{\gamma\gamma} < 1$~$\mathrm{GeV/c}$) events. Three statistical methods were applied to assess the significance of the enhancement. All methods yield a statistical significance above 3.7$\sigma$. The spectral fitting indicates a mass reduction of $57.5^{+~5.7}_{-27.8}$~$\mathrm{MeV/c}^{2}$.
Our result provides the first evidence for an in-medium spectral modification of the $\eta^{\prime}$ meson obtained through a direct measurement of its invariant mass.
\end{abstract}


\maketitle



{\it Introduction.}--
Understanding the behavior of mesons within the nuclear medium offers vital insights into the non-perturbative aspects of Quantum Chromodynamics (QCD) \cite{REVIEW01,HAYANO2010,REVIEW02}. 
In particular, the low-lying pseudoscalar nonet provides information deeply connected to the underlying symmetries of QCD. 
Under $U(3)_L \times U(3)_R$ chiral symmetry that can be organized into $SU(3)_V \times SU(3)_A \times U(1)_V\times U(1)_A$, the spontaneous breaking of $SU(3)_A$ symmetry 
gives rise to the octet pseudoscalar Nambu-Goldstone (NG) bosons (3$\pi$'s, 4$K$'s, and $\eta$) \cite{NAMBU1960-2,NAMBU1960,NJL}.
In contrast, the  $U(1)_A$ symmetry plays a crucial role in shaping the mass spectrum, especially of the $\eta'(958)$ meson that is known as a combination of the singlet $\eta$ ($\eta_0$) as a main component and the octet $\eta$ ($\eta_8$). 
The $\eta_0$ cannot serve as an NG boson due to the presence of the axial $U(1)$ anomaly.  
The $\eta'$ meson, therefore, has a significantly larger mass compared with the mass of the NG bosons in vacuum \cite{WEINBERG1975,WITTEN1979,VENEZIANO1979}. 

The interior of a nucleus exhibits a highly dense environment of approximately $2\times 10^{14}$~$\mathrm{g/cm^3}$, resembling the universe shortly after the Big Bang, where chiral symmetry may partially be restored \cite{HATSUDALEE1992,HAYANO2010}. 
It is anticipated that the mass of the $\eta'$ meson ($\sim\eta_0$) decreases, similar to a typical $q\bar{q}$ meson in the nuclear medium. 
Simultaneously, the axial $U(1)$ anomaly may also be less significant if the in-medium $\eta^\prime$ mass reduction happens.
In this context, theoretical calculations that consider both spontaneous breaking of chiral symmetry and axial $U(1)$ anomaly predict the in-medium $\eta'$ mass deviation $\Delta m_{\eta'}$ of approximately $-150$ \cite{COSTA2003.171,COSTA2005.71,PhysRevC.74.045203} and $-(80$--$100)$ \cite{PhysRevC.88.064906} $\mathrm{MeV/c^2}$ at the normal nuclear density, based on the Nambu--Jona-Lasinio (NJL) and linear $\sigma$ models, respectively.
Note that there is another prediction by the Quark Meson Coupling (QMC) model \cite{SAITO20071}, giving $\Delta m_{\eta'}\sim -60$~$\mathrm{MeV/c^2}$ \cite{BASS2006368} according to the mixing angle of $\eta_0$ and $\eta_8$.

However, it remains an open question whether the suppression of anomaly effects accompanies the restoration of chiral symmetry, although the interplay between them is being suggested theoretically \cite{10.1093/ptep/ptab084,LEE1996,JIDO2012}. 
Examining the spectral function of $\eta'$ in the nuclear medium opens a new avenue for understanding the impacts of chiral symmetry restoration and the axial $U(1)$ anomaly under dense conditions.
For this purpose, the present work aims to measure the invariant mass of $\eta'$ mesons within a nucleus through the $\eta'\to\gamma\gamma$ channel.
This channel has an advantage because the $\gamma\gamma$ invariant mass directly reflects the in-medium $\eta'$ mass, 
giving clearer signals with no final state interactions.
There should also be no influence on the two $\gamma$'s four-momenta by energy loss.

{\it Previous experimental studies.}--
Although no evidence for an in-medium modification of the $\eta^{\prime}$ mass spectrum has been reported to date, several experimental studies have investigated the $\eta^{\prime}$ mass in nuclei using indirect methods.
The CBELSA/TAPS collaboration measured the depth of the $\eta^{\prime}$–nucleus potential, $V_{0}\sim -40~\mathrm{MeV}$ \cite{NANOVA2013,NANOVA2016,NANOVA2018}, which corresponds to the mass deviation at normal nuclear density.
The COSY-11 collaboration extracted the scattering length of $\etap p$ interaction \cite{PhysRevLett.113.062004}.
They claimed that the value is consistent with the prediction by the QMC model, while it disfavors the expectation by the NJL and linear $\sigma$ models.
Furthermore, the $\eta$-PRiME/Super-FRS collaboration \cite{YKTANAKA2018} and the LEPS2/BGOegg collaboration \cite{TOMIDA2020} searched for $\eta^{\prime}$-nucleus bound states, which could be formed if the $\eta^{\prime}$–nucleus system offers a deeply attractive potential with small absorption.
Their studies have not revealed any bound states, indicating that a large mass reduction exceeding $100~\mathrm{MeV/c^2}$ is disfavored.

On the other hand, an analysis using the RHIC data of Au+Au collisions has claimed an $\etap$ mass reduction larger than $200~\mathrm{MeV/c^2}$ based on the two-pion Bose-Einstein correlation functions~\cite{HIC2010}.
However, a similar analysis using the CERN SPS data has indicated no significant mass modification~\cite{HIC2011}.

{\it Experiment.}--
The experiment was conducted using a 20 mm thick carbon target on the SPring-8 LEPS2 beamline \cite{MURAMATSU2014184}, which provided a tagged photon beam with the intensity of a few times $10^6$/s in the energy range of $1.3$--$2.4$~$\mathrm{GeV}$.
This beam was produced by laser Compton scattering,
being clean with small low-energy components.
A high-resolution electromagnetic calorimeter, BGOegg \cite{ISHIKAWA2016109}, was used to measure the energies and angles of two photons.
The precise measurement of them is crucial to find any anomalous shape in the $\gamma\gamma$ invariant mass spectrum.
The BGOegg calorimeter is an egg-shaped assembly of 1320 BGO crystals, covering the polar angles from 24$^\circ$ to 144$^\circ$ with a good segmentation of $4$--$6^\circ$ in both polar and azimuthal directions.
Each crystal has a sufficient thickness of 220 mm, corresponding to 20 radiation lengths.
It features a self-supporting structure that eliminates the insensitive area between adjacent crystals.
This design enables BGOegg to achieve an energy resolution of 1.38\% for 1 GeV photons, which is among the highest globally.
A charged particle was also detected by this calorimeter, together with a hit signal at the corresponding inner plastic scintillator.
In addition, a planar drift chamber covered polar angles less than $22^\circ$ for the detection of charged particles.
Experimental details are given in Refs. \cite{PhysRevC.106.035201,PhysRevC.100.055202,ISHIKAWA2016109}.

{\it Analysis.}--
The $\eta^{\prime}\to\gamma\gamma$ events were selected under the condition that exactly two photons were detected by the BGOegg calorimeter.
The invariant mass spectrum was reconstructed from the four-momenta of these photons.
At most one charged particle was allowed in the final state to select the quasi-free reactions $\gamma N\to\eta^{\prime}N$, where $N$ denotes a proton or a neutron. No further constraints on the recoiling proton or neutron were imposed in the analysis.
Additional criteria on the electromagnetic shower shape and hit timing of the detected $\gamma$'s were required to reduce background events arising from the misidentification of neutrons, cosmic rays, and particles produced upstream of the detector setup.
The $\gamma$ selection criteria on the electromagnetic shower shape were optimized using a Monte Carlo (MC) simulation based on GEANT4 \cite{GEANT4}.
These criteria have been confirmed not to introduce a bias in the $\gamma\gamma$ invariant mass distribution.

The energy calibration of the BGOegg calorimeter was performed using the $\pi^{0}$ and $\eta$ mass peaks in the $\gamma\gamma$ invariant mass distributions.
Energy leakage from a $\gamma$ cluster reconstructed by connecting hit crystals was evaluated for its correction as a function of energy and hit position using the MC simulation.
The reconstructed meson masses were confirmed to be independent of momentum and emission angle over a wide range, validating the quality of the calibration.

\begin{figure}[t]
  \centering
  \includegraphics[width=8.1cm]{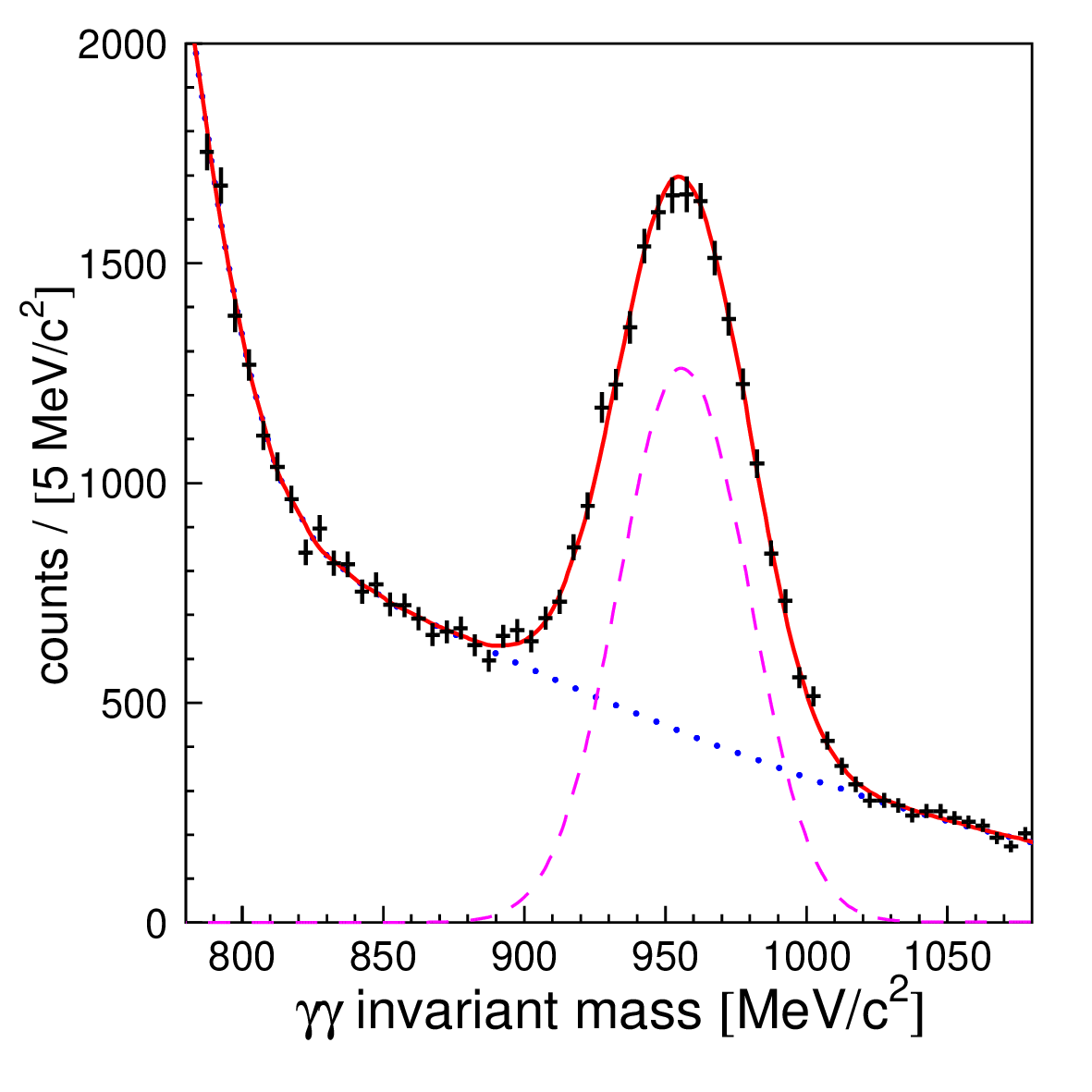}
  \caption{The $\gamma\gamma$ invariant mass spectrum in the high-momentum sample with the fit result (red-solid line). The magenta-dashed and blue-dotted lines represent the template shape for $\etap$ decays in vacuum ($F_\mathrm{QF}$) and the sum of background functions ($F_\mathrm{M}+F_\omega$), respectively.}
  \label{highpfit}
\end{figure}
\begin{figure}[t]
  \centering
  \includegraphics[width=8.1cm]{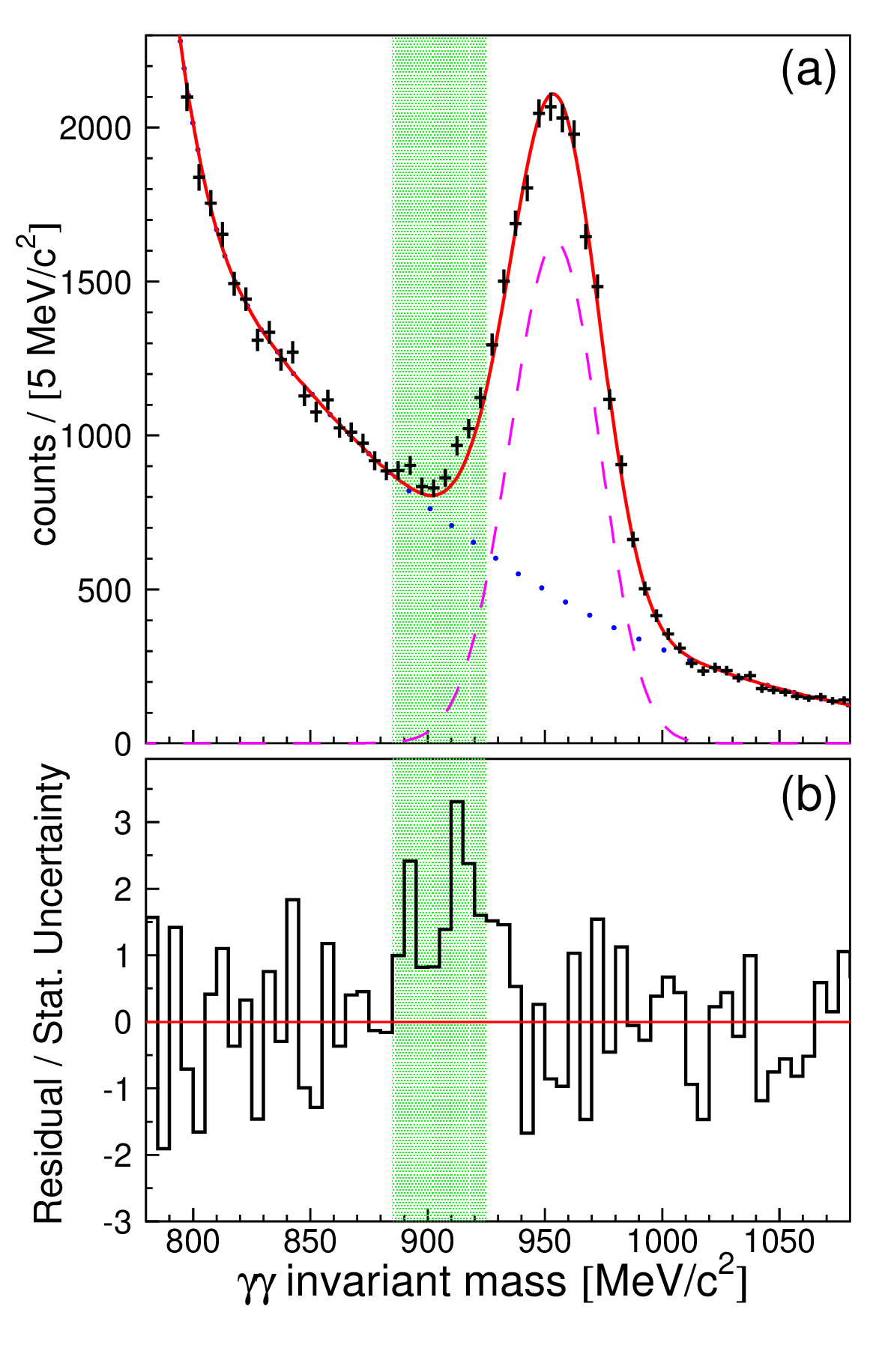}
  \caption{(a) The $\gamma\gamma$ invariant mass spectrum in the low-momentum sample with the fit result. The notation of lines is the same as that in Fig.~\ref{highpfit}. The green hatched area indicates the mass region excluded from the fit. (b) The residuals divided by statistical uncertainties in each bin for the fit in panel (a).}
  \label{bgonlyfit}
\end{figure}

{\it Spectral fit.}--
As shown in Fig.~\ref{highpfit} and Fig.~\ref{bgonlyfit}(a), the line shape of the $\etap$ meson mass in the measured $\gamma\gamma$ invariant mass distribution was investigated by fitting the spectral functions that described the relevant processes realistically.
The obtained mass distribution is explained by the peak of quasi-free $\eta^{\prime}$ production with the $\gamma\gamma$ decay in vacuum and the background spectra that originate from the processes to produce multi-mesons (e.g., $\pi^0\pi^0$, $\pi^0\eta$) and the higher mass tail contribution of $\omega$ photoproduction.
If the line shape cannot be reproduced by these known processes, the data suggest a spectral change possibly due to the in-medium $\etap$ mass reduction.
The fitting function for the quasi-free $\eta^\prime$ peak ($F_{\mathrm{QF}}$) was prepared as a template histogram by generating the process $\gamma p \to \eta^\prime p \to \gamma \gamma p$ in the MC simulation and applying the same event reconstruction and selection as for the experimental data. The energy resolution as well as the detector geometry were precisely adjusted in the simulation so as to thoroughly reproduce the mass resolutions of mesons in the real data.
The typical mass resolutions~($\sigma$~of the peaks) for $\pi^0$, $\eta$, and $\eta^\prime$ mesons are $6.7$, $14.4$, and $20.7$~$\mathrm{MeV/c^2}$, respectively.

The multi-meson production events yielding four photons in the final state contribute as a background when two photons are undetected due to the calorimeter acceptance.
This background distribution smoothly decreases as the $\gamma\gamma$ invariant mass increases, not showing any particular structure over a wide range.
Its smooth shape was verified by the MC simulations of not only non-resonant $\pi^0\pi^0 N$ and $\pi^0\eta N$ processes but also $\pi^0\Delta$, $\pi^0 N^*$, and $a_0(980)N$ reactions.
It was confirmed that any arbitrary mixture of the above processes was expressed by a simple function $F_{\mathrm{M}}(x)=\exp(p_{0}+p_{1}x+p_{2}x^{2}+p_{3}x^{3}$), where $p_{0}$--$p_{3}$ are free parameters.
Other multi-meson production processes such as $\pi^{0}\pi^{0}\pi^{0} N$ and $\pi^{0}\pi^{0}\eta N$ were
negligible after the event selection.
Photoproduction of the $\omega$ meson decaying into $\pi^0\gamma$ less significantly contributes as a background when one photon in $\pi^0\to\gamma\gamma$ is missing.
Its contribution to the $\gamma \gamma$ invariant mass spectrum was estimated as a template histogram ($F_\omega$) by a realistic MC simulation, as done for the quasi-free $\eta^\prime$ peak.

In the following spectral fits, the sum of the three components ($F_{\mathrm{QF}}$, $F_{\mathrm{M}}$, and $F_\omega$) was treated as the fitting function expressing the $\gamma \gamma$ invariant mass spectrum for the case without a medium modification signal.
The fit was performed with seven free parameters: the amplitudes of $F_{\mathrm{QF}}$ and $F_{\omega}$, $p_{0},\cdots,p_{3}$, and a mass shift parameter $\mu_{m}$, which was common for the three components.
The parameter $\mu_{m}$ was introduced to adjust for a small mass deviation which possibly exists due to the individual calibrations for the real and MC data.

{\it Analysis sample.}--
The probability that $\eta^{\prime}$ decays inside a nucleus is expected to increase as the $\eta^{\prime}$ momentum decreases, owing to its shorter decay length in the laboratory frame.
Therefore, low-momentum events were used for the line-shape analysis, while high-momentum events served as a reference sample.
Prior to the line-shape analysis, the momentum cut point $P_{\mathrm{cut}}$ was optimized to maximize the significance of a possible mass reduction signal arising from in-medium decays, $N_{\mathrm{in}}/\sqrt{N_{\mathrm{oth}}}$, supposing the statistics of the present experiment for the low-momentum $\gamma \gamma$ sample ($P_{\gamma \gamma} < P_{\mathrm{cut}}$).
Here, $N_{\mathrm{in}}$ and $N_{\mathrm{oth}}$ represent the expected numbers of in-medium $\eta^\prime$ decays and the other events originating from the above known processes, respectively.
These numbers were estimated as a function of $P_\mathrm{cut}$ using MC simulations, so that the determination of $P_\mathrm{cut}$ was blinded to the signal significance in the line-shape analysis of the experimental data.
Finally, the sample of $\gamma \gamma$ events was divided into two regions: $P_{\gamma \gamma} < 1.0$ and $1.0 < P_{\gamma \gamma} < 1.5$~$\mathrm{GeV/c}$ (corresponding to $\beta\gamma < 1.04$ and $1.04 < \beta\gamma < 1.57$ for the $\etap$ meson), called the low- and high-momentum samples, respectively.
The statistics of the low- and high-momentum samples reached $64$K and $50$K events, respectively, in the mass range of $780$--$1080$~$\mathrm{MeV/c^2}$, which was used for the spectral fit described below.

{\it Fit results by known processes.}--
To examine the existence of spectral change, the sum of the spectral functions prepared for the known processes was fitted to the $\gamma \gamma$ invariant mass spectra.
The spectrum in the high-momentum sample was well reproduced by this fit, giving $\chi^2 = 48.7$ for $ndf = 54$ (Fig.~\ref{highpfit}), whereas the same fit to the low-momentum sample resulted in $\chi^2 = 69.2$, showing a clear excess over the fitted model shape only in the lower tail of the $\eta^\prime$ mass peak.
Thus, a modified fit was performed by excluding the mass range, $885$--$925$~$\mathrm{MeV/c^2}$, corresponding to the observed excess.
Figure~\ref{bgonlyfit}(a) and (b) show this spectral fit for the low-momentum sample and a residual distribution after subtracting the fit result, respectively.
A clear enhancement was seen around $910$~$\mathrm{MeV/c^2}$ with the significance of $4.7 \sigma$, which was evaluated by dividing the residual sum by the square root of the total number of events in the excluded region.
As shown in End Matter, this significance was reliable even if the boundaries of the excluded mass range were varied.
The observed enhancement simply provides evidence for the spectral change possibly due to in-medium $\eta^\prime$ mass reduction.

{\it Signal simulation.}--
To pin down the observed spectral change, how the signal of in-medium $\eta^\prime$ mass reduction could appear in the $\gamma \gamma$ invariant mass distribution was investigated using a simple model in the realistic MC simulation.
$\eta^\prime$ mesons were first generated through the quasi-free $\gamma p \to \eta^\prime p$ process inside the $^{12}\mathrm{C}$ nucleus target by using the known differential cross sections \cite{CLAS2009}.
The subsequent propagation of $\eta^\prime$ inside a $^{11}\mathrm{B}$ nucleus was then simulated with continuous fine steps while requiring the energy conservation for the $\eta^\prime$ motion until its decay.
The decay point was stochastically determined by calculating the decay probability at each step based on the $\eta^\prime$ lifetime obtained from its total decay width, described below.
The $\gamma \gamma$ decay was caused according to the branching fraction, calculated by the ratio of partial to total decay widths at the decay point.

The medium modification was incorporated by considering the density-dependent $\eta^\prime$ mass $m$ and total decay width $\Gamma^{\mathrm{tot}}$ with the assumption of linear dependence on the nuclear density $\rho$ \cite{MUTO2007}:
\begin{eqnarray}
m(\rho) &=& m_{0}\left(1-k_{1}\frac{\rho}{\rho_{0}}\right)\label{eq:1}\\
\Gamma^{\mathrm{tot}}(\rho) &=& \Gamma_{0}^{\mathrm{tot}}\left(1+k_{2}\frac{\rho}{\rho_{0}}\right),
\end{eqnarray} 
where $m_{0}=957.78$~$\mathrm{MeV/c^2}$ and $\Gamma_{0}^{\mathrm{tot}}=0.188$~$\mathrm{MeV}$ \cite{pdg} indicate the values in vacuum, and $\rho_0$ represents the normal nuclear density.
In our simulation, $k_2$ was fixed to $105$, corresponding to a width broadening of $20$~$\mathrm{MeV}$ measured by the CBELSA/TAPS collaboration \cite{NANOVA2012600}, and the partial $\gamma \gamma$ decay width was unchanged from the value in vacuum.
The local nuclear density $\rho$ at each propagation step was estimated with the Woods-Saxon distribution
$\rho(r)\propto [1+\exp\left\{(r-R)/a\right\}]^{-1}$, where $r$ was the distance from the center of a nucleus.
The Woods-Saxon parameters were taken as $R = 2.78$~$\mathrm{fm}$ and $a = 0.65$~$\mathrm{fm}$ for the radius and surface thickness of a $^{11}\mathrm{B}$ nucleus, respectively \cite{GLOVER1980}.
This density distribution was also utilized to determine the position-dependent $\eta^\prime$ production probability inside a $^{12}\mathrm{C}$ nucleus with $R = 2.86$~$\mathrm{fm}$.
In the simulation for a certain $k_1$, in-medium $\eta^\prime$ decays into $\gamma \gamma$ were generated using $m(\rho)$ at the decay points.
The generated $\gamma \gamma$ pairs were further simulated with the realistic detector response to get a template histogram as the signal function for a spectral fit.
The signal function was made for the decays at $r<R+4a$.

{\it Fit results with a signal function.}--
\begin{figure}[b]
    \centering
    \includegraphics[width=8.2cm]{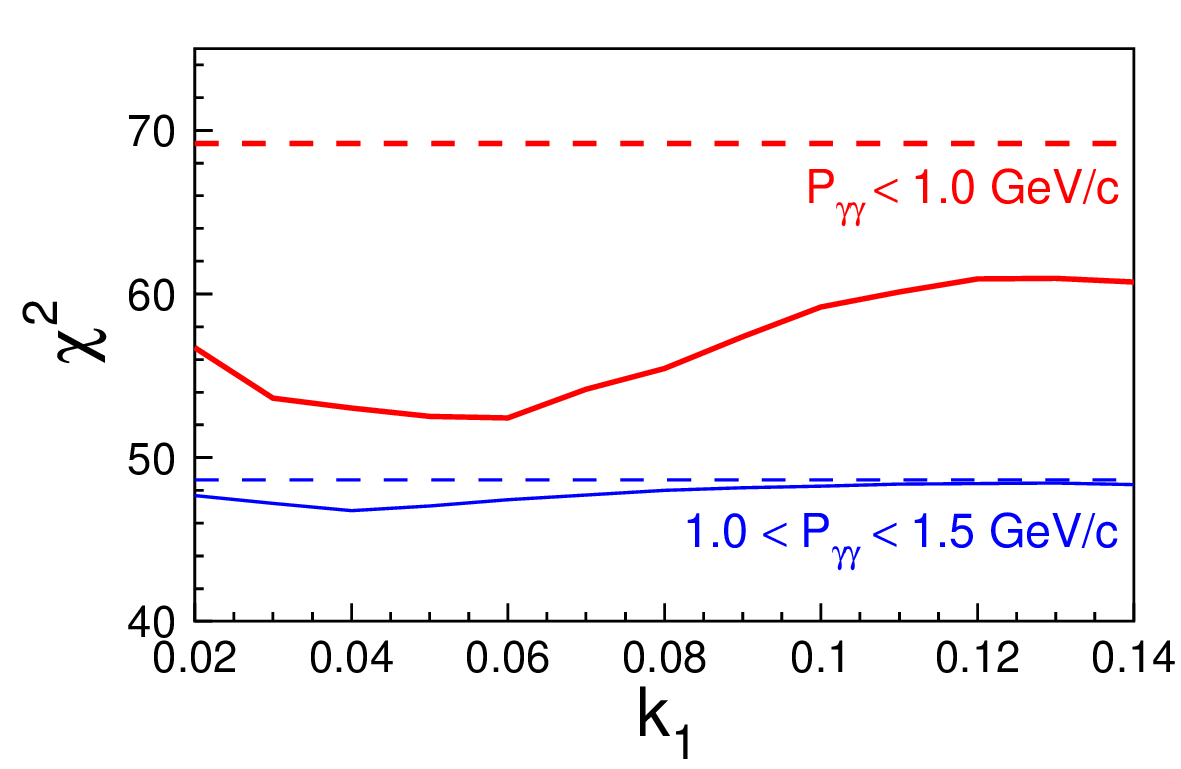}
    \caption{The $k_1$ dependence of $\chi^2$'s for the fits with simulated signal functions (solid lines). Results in both lower and higher momentum samples are shown by red-thick and blue-thin lines, respectively. The dashed lines indicate $\chi^2$'s for the fits without the signal functions.}
    \label{chi2difftest}
\end{figure}
As a complementary inspection, all the spectral functions including the signal template histogram with a free normalization parameter were simultaneously fitted to the $\gamma \gamma$ invariant mass distribution using the same fitting range as the initial method but without any exclusion.
This alternative procedure has merits to discuss the spectral change more deeply.
To find the most probable model, the fits were repeated while changing the signal function whose shape was simulated with the $0.01$-step variation of $k_1$ from $0.02$ to $0.14$.
At each $k_1$, signal significance was calculated by the $\chi^2$ difference test that compared the models including and omitting the signal function.
For the low-momentum sample, the best fit was obtained at $k_1=0.06$, providing $\chi^2 = 52.4$ for $ndf=53$ (see Fig.\ref{fig:5} in End Matter).
The significance was evaluated to be $4.1\sigma$ based on the $\chi^2$ difference $\Delta \chi^2 = 16.8$ with $\Delta ndf=1$.
The $\Delta\chi^2$ decreases as $k_1$ is apart from $0.06$ as shown in Fig.~\ref{chi2difftest}, so that the most probable $k_1$ is uniquely determined together with the uncertainties of $-0.029$ and $+0.006$, which are estimated from the $k_1$ values at $\Delta \chi^2 = 16.8 - 1$.
Even if $k_1$ is conservatively considered as a free parameter, the signal significance is still $3.7 \sigma$ based on $\Delta \chi^2 = 16.8$ and $\Delta ndf = 2$.
In the fits for the high-momentum sample, the significances do not exceed $1.4 \sigma$ over the same $k_1$ range, suggesting no need to include a signal component into the fit model.

In addition, the above alternative procedure was modified by replacing the simulated signal histogram with a Gaussian function in order to examine the validity of results using a different signal shape assumption.
The $\chi^2$ difference tests using the Gaussian function with free normalization and $\sigma$ parameters ($\Delta ndf = 2$) were performed while its mean ($\mu$) was scanned from $890$ to $935$~$\mathrm{MeV/c^2}$ with a $5$~$\mathrm{MeV/c^2}$ step.
For the low-momentum sample, the maximum significance of $4.0 \sigma$ ($\Delta \chi^2 = 19.4$) was observed at $\mu = 925$~$\mathrm{MeV/c^2}$ with $\sigma = 21.4$~$\mathrm{MeV/c^2}$.
By the same discussion as the fit with a signal template, the conservative significance for the case of $\Delta ndf = 3$ was calculated to be $3.7 \sigma$.
The same analysis for the high-momentum sample returned the significance less than $1.3 \sigma$ at any $\mu$.
This modified method supports the results of the other two spectral fits.

{\it Reference analyses.}--
Because the spectral enhancement is seen in the lower tail of the $\eta^\prime$ mass peak, the validity of the fitting peak shape estimated for quasi-free $\etap$ decaying in vacuum is essential.
This shape was obtained from the MC simulation that should reproduce the experimental data in terms of $\gamma \gamma$ invariant mass resolution and energy leakage from a reconstructed $\gamma$ cluster.
Although such reproducibility was confirmed by showing no signal indication in the high-momentum sample, further examination was performed for the $\gamma \gamma$ invariant mass spectrum around the $\eta$ peak.
Ten examination samples with statistics similar to the low-momentum $\etap$ sample were selected by requiring $P_{\gamma \gamma} < 0.6$~$\mathrm{GeV/c}$ to focus on the same $\beta\gamma=P_{\gamma \gamma} / M$ range as the $\etap$ case, where $M$ denotes the mass of either meson.
Spectral fits in the range of $460$--$640$~$\mathrm{MeV/c^2}$ were performed using a Gaussian signal component, a third-order polynomial background function, and the template peak shape simulated for the $\eta$ decays in vacuum.
The mean $\mu$ of a signal component was scanned from $490$ to $530$~$\mathrm{MeV/c^2}$ for $\chi^2$ difference tests, and no need to include the signal component in the fitting model was found because the signal significances were lower than $0.9\sigma$ over the inspected values of $\mu$ in the ten samples.

In addition, data collected with a liquid hydrogen (LH$_2$) target \cite{PhysRevC.106.035201,PhysRevC.100.055202} was examined by selecting the low-momentum $\eta^\prime$ sample with the same conditions as the carbon target case.
To analyze this sample, the template shape for $\eta^\prime$ decays in vacuum was prepared by incorporating the target difference into the simulation.
The $\gamma \gamma$ invariant mass spectrum was well described only by this template distribution and background functions with $\chi^2 = 23.1$ for $ndf = 24$.
In the $\chi^2$ difference tests with the simulated signal functions, the signal significances were less than $1.1 \sigma$ over the inspected $k_1$ range.

\begin{figure}[b]
 \centering
 \includegraphics[width=8.2cm]{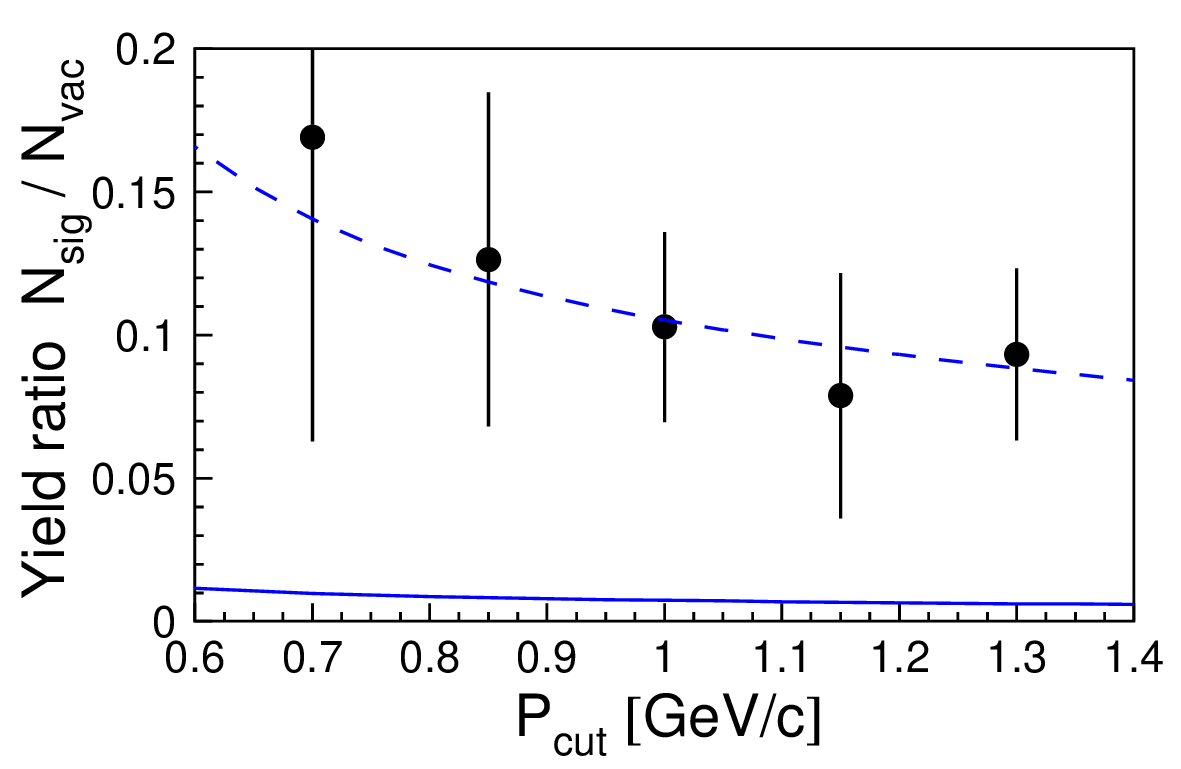}
 \caption{The yield ratio of the fitted signal component to the $\eta^\prime$ decays in vacuum as a function of $P_{\mathrm{cut}}$ to select a low-momentum sample (black filled points). The vertical bars represent statistical uncertainties. The solid and dashed lines indicate the ratio calculated in the signal simulation and the scaled ratio described in the text, respectively.}
 \label{pdep}
\end{figure}

{\it Discussions.}--
For the carbon target data, the $P_{\gamma \gamma}$ dependence of the observed enhancement was examined more precisely using the fitting model with the signal function simulated with $k_1=0.06$.
Figure~\ref{pdep} shows the yield ratio of the signal component $N_{\mathrm{sig}}$ to the $\eta^\prime$ decays in vacuum $N_{\mathrm{vac}}$ for the different requirements of $P_{\gamma \gamma} < P_{\mathrm{cut}}$.
Although the statistical uncertainty is large, the measured ratio increases by selecting lower $P_{\gamma \gamma}$ events.
The $P_{\gamma \gamma}$ dependence is similar between the measurement and the expectation from the simulation of in-medium $\eta^\prime$ mass reduction, as shown by the dashed line.
This line was obtained by a fit of the simulation result (solid line) with a scale factor of $14.3$.
The need for a scale factor implies the possibility of the partial $\gamma \gamma$ decay width variation inside a nucleus \cite{MUTO2007}.
Such a change may be partly explained by the medium modification of the $\eta^\prime$ wave function \cite{10.1093/ptep/ptu023}.

In the spectral fit with the simulated signal function, the total decay width parameter $k_2$ was fixed to $105$.
Even if $k_2$ was varied from $50$ to $500$, the most probable $k_1$ value remained unchanged while showing a large uncertainty for $k_2$.
From Eq.~\ref{eq:1}, the measured $k_1$ value corresponds to the $\etap$ mass reduction of $57.5^{+~5.7}_{-27.8}$~$\mathrm{MeV/c^2}$ at the normal nuclear density.
It matches a lower range of the current theoretical predictions.
In addition, it does not contradict the previous results of the search for $\etap$ mesic nuclei \cite{YKTANAKA2018,TOMIDA2020}.

{\it Conclusions.}--
The present analysis searched for the in-medium mass spectral change of the $\eta^\prime$ meson, which was photoproduced using a carbon target and was detected through the $\gamma \gamma$ decay channel.
In the fit using the spectra of known processes, an enhancement was observed in the lower tail of the $\eta^\prime$ mass peak only when requiring $P_{\gamma \gamma} < 1$~$\mathrm{GeV/c}$.
Its significance was $4.7 \sigma$ by a fit excluding the excessive region.
In another spectral fit together with a simulated signal function, the best fit was obtained at $k_1 = 0.06^{+0.006}_{-0.029}$, indicating the significance of $4.1 \sigma$ (conservatively $3.7 \sigma$) based on the $\chi^2$ difference test.
The measured $k_1$ value corresponds to a mass reduction of $57.5^{+~5.7}_{-27.8}$~$\mathrm{MeV}/c^{2}$ at the normal nuclear density.
The present result is the first evidence of the spectral change in the direct measurement of the in-medium $\eta^\prime$ mass.\\

{\it Acknowledgments.}--
The experiment was performed at the BL31LEP of SPring-8 with the approval of the Japan Synchrotron Radiation Institute (JASRI) as a contract beamline (Proposal No.~BL31LEP/6101).
The authors gratefully acknowledge to the support of the staff at SPring-8 for providing excellent experimental conditions.
This research was supported in part by the Ministry of Education, Culture, Sports, Science and Technology of Japan, JSPS KAKENHI Grant Nos.~19002003, 24244022, 21H04986, 20K03984, and the National Science and Technology Council of the Republic of China (Taiwan).

\bibliographystyle{apsrev4-2} 
\bibliography{refs}

\onecolumngrid
\section*{End Matter}
\twocolumngrid
This section describes the additional materials that support discussions in the spectral fits for the low-momentum $\etap$ sample of the carbon target data and the two reference analyses conducted for the validity test.

{\it Spectral fit with an excluded mass range.}--
For the low-momentum sample, a spectral fit using only the functions that described known processes was performed by excluding the mass range of $885$--$925$~$\mathrm{MeV/c^2}$ by default.
As written in the main text, this fit results in $N_{in} / \sqrt{N_{oth}} = 4.7 \sigma$ as the significance of the observed enhancement in the excluded region.
In the present work, the robustness of this result was examined by varying the boundaries of the excluded mass range by $\pm 5$~$\mathrm{MeV/c^2}$.
The middle column of Table~\ref{tab:6-1} shows the significances obtained in the cases of the boundary variation, and exhibits $4.1$--$4.8 \sigma$ supporting the default result.
In contrast, the same test using the high-momentum sample returned low significances as shown in the right column of Table~\ref{tab:6-1}.
\begin{table}[b]
    \begin{center}
        \caption{Significances in low-momentum ($P_{\gamma\gamma}< 1.0$) and high-momentum ($1.0 < P_{\gamma\gamma} < 1.5$~$\mathrm{GeV/c}$) samples obtained by the fit with excluded mass region.}
        \label{tab:6-1}
        \resizebox{\linewidth}{!}{
        \begin{tabular}{ccccc} \hline\hline
            Excluded region & & Significance ($\sigma$) & & Significance ($\sigma$) \\ 
            ($\mathrm{MeV/c^2}$) & & for low-$P_{\gamma \gamma}$ & & for high-$P_{\gamma \gamma}$ \\ \hline
            $880\leq M_{\gamma\gamma}<930$ & & $4.8$ & &  $1.4$    \\
            $885\leq M_{\gamma\gamma}<930$ & & $4.8$ & &  $1.6$    \\
            $890\leq M_{\gamma\gamma}<930$ & & $4.7$ & &  $2.3$    \\
            $880\leq M_{\gamma\gamma}<925$ & & $4.6$ & &  $0.3$    \\
            $885\leq M_{\gamma\gamma}<925$ & & $4.7$ & &  $0.5$    \\
            $890\leq M_{\gamma\gamma}<925$ & & $4.6$ & &  $1.2$    \\
            $880\leq M_{\gamma\gamma}<920$ & & $4.1$ & & $-0.3$
            \\
            $885\leq M_{\gamma\gamma}<920$ & & $4.4$ & &  $0.3$  
            \\
            $890\leq M_{\gamma\gamma}<920$ & & $4.4$ & &  $1.1$
            \\
            \hline\hline
        \end{tabular}
        }
    \end{center}
\end{table}

\begin{figure}[t]
  \centering
  \includegraphics[width=8.1cm]{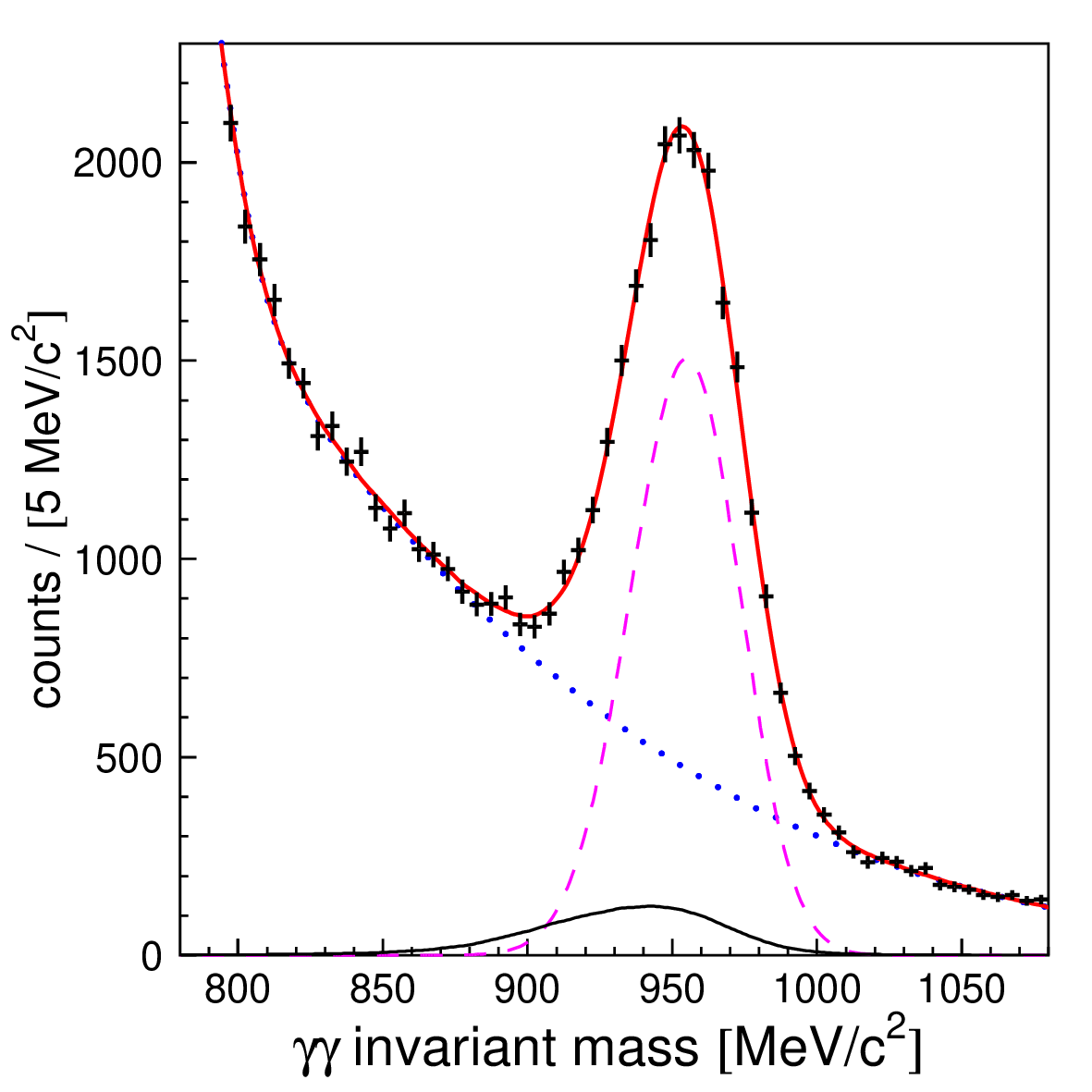}
  \caption{The $\gamma\gamma$ invariant mass spectrum in the low-momentum sample with the fit result (red-thick-solid line). The signal component for the $k_1$ value of 0.06 (black-thin-solid line) was simultaneously fitted with the template shape for $\etap$ decays in vacuum (magenta-dashed line) and the sum of background functions (blue-dotted line).}
  \label{fig:5}
\end{figure}

{\it Spectral fit with a simulated signal template.}--
Secondly, a spectral fit was done using the model that summed up $F_{QF}$, $F_{M}$, $F_\omega$, and a template histogram simulated for the in-medium $\eta^\prime$ mass reduction signal.
The best fit for the low-momentum sample was obtained with the significance of $4.1 \sigma$ when fitting the signal template simulated by the mass reduction parameter $k_1 = 0.06$.
Figure~\ref{fig:5} shows the result of this best fit, where the black solid curve indicates the signal component.
Although the $k_1$ value of $0.06$ means the mass reduction of $57.5$~$\mathrm{MeV/c^2}$ at the normal nuclear density, the signal distribution is widely spread because the signal simulation has been done in a large volume of $r<R+4a$, which contains the positions with low nuclear densities following the Woods-Saxon distribution.
Note that the rate of $\eta^\prime$ decaying within the nuclear radius $R$ amounts to $25\%$ in the simulated signal component.

\begin{figure}[b]
 \centering
 \includegraphics[width=4.2cm]{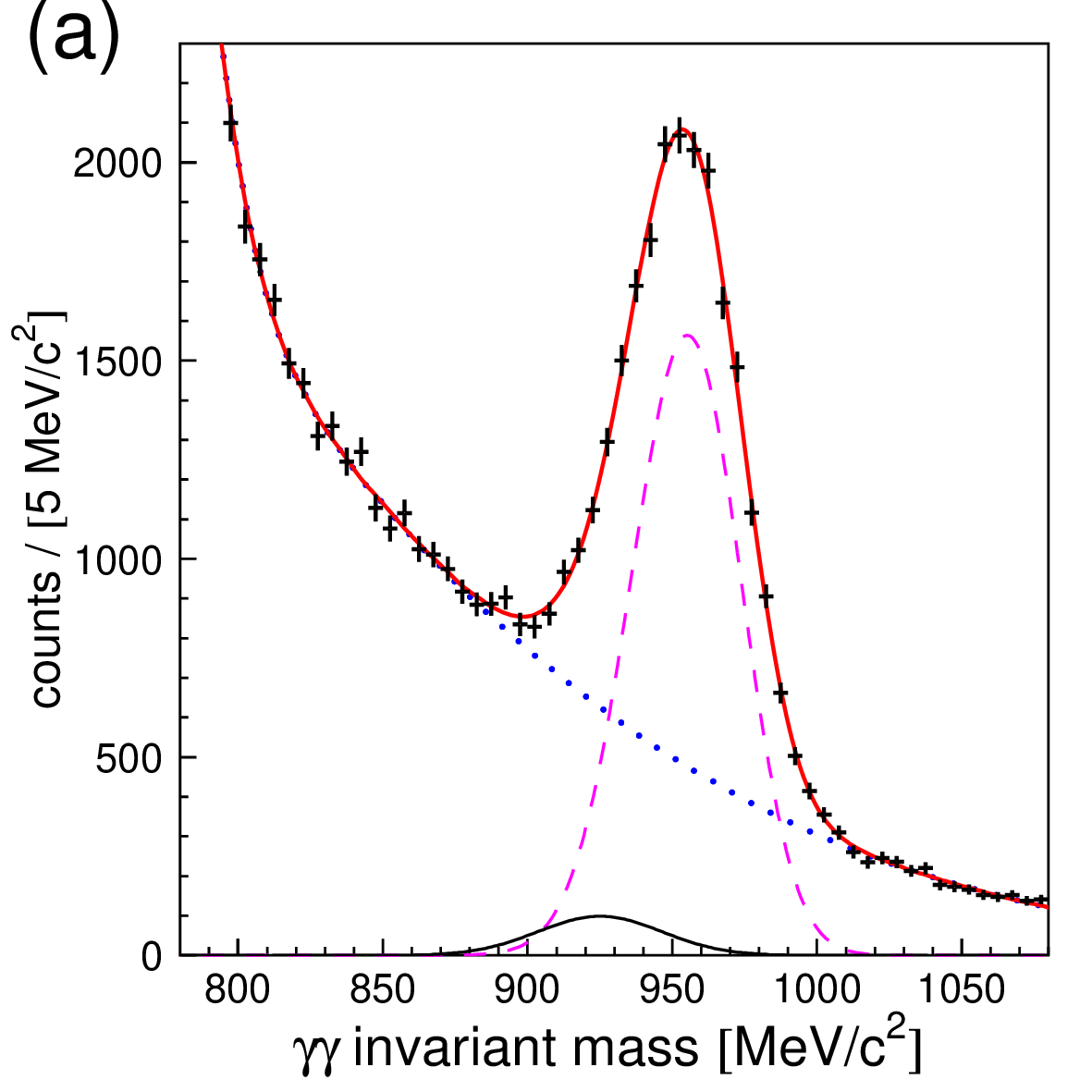}
 \includegraphics[width=4.2cm]{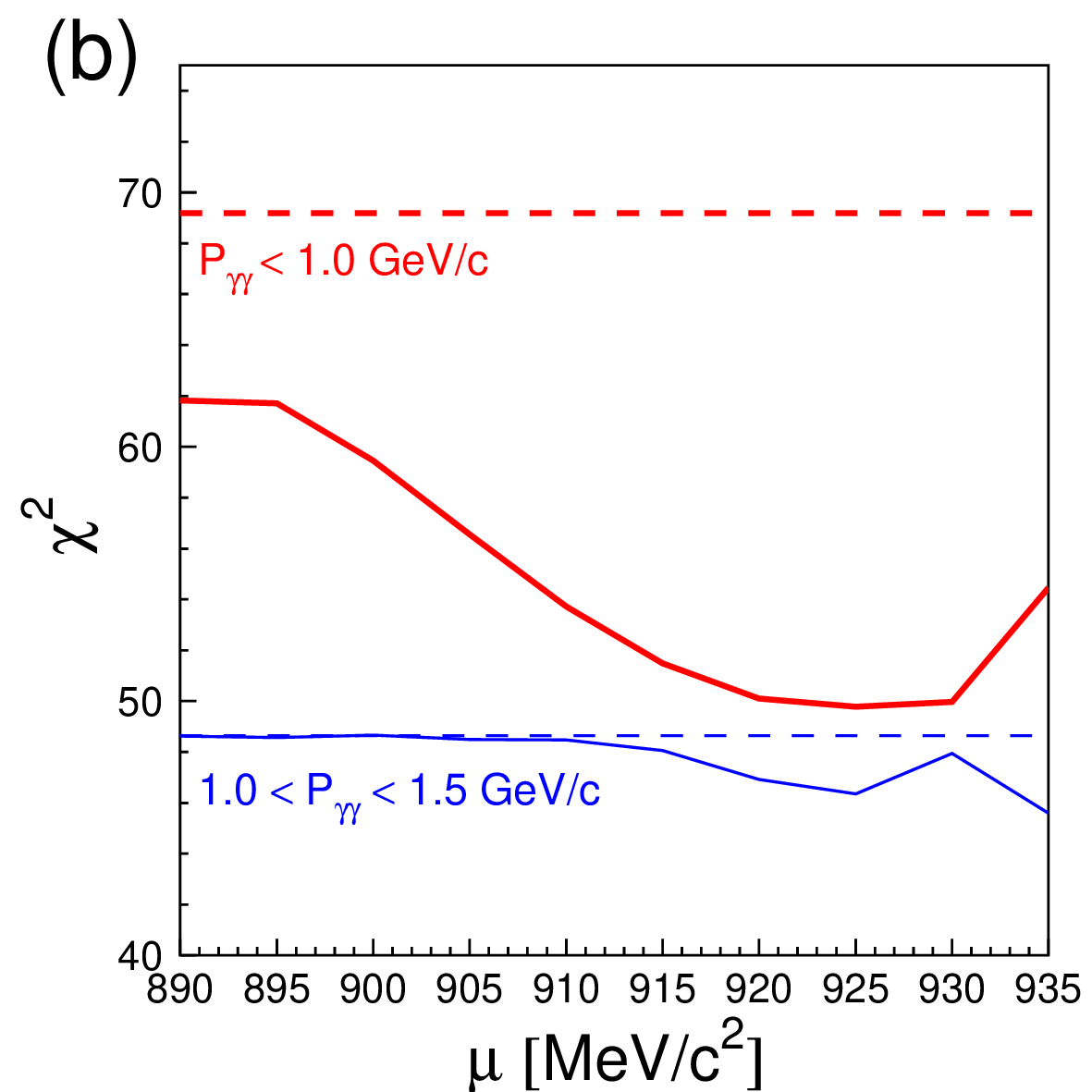}
 \caption{(a) The $\gamma\gamma$ invariant mass spectrum in the low-momentum sample with the best fit result for the case including a Gaussian signal function (black-thin-solid line). The notation of lines is the same as that in Fig.~\ref{fig:5}. (b) The $\mu$ dependence of $\chi^2$'s for the fits with Gaussian signal functions. The notation of lines is the same as that in Fig.~\ref{chi2difftest} in the main text.}
 \label{fig:6}
\end{figure}

\begin{figure*}
 \begin{tabular}{cc}
  \begin{minipage}[]{\textwidth}
   \centering
    \begin{minipage}[]{0.48\textwidth}
     \includegraphics[width=4.2cm]{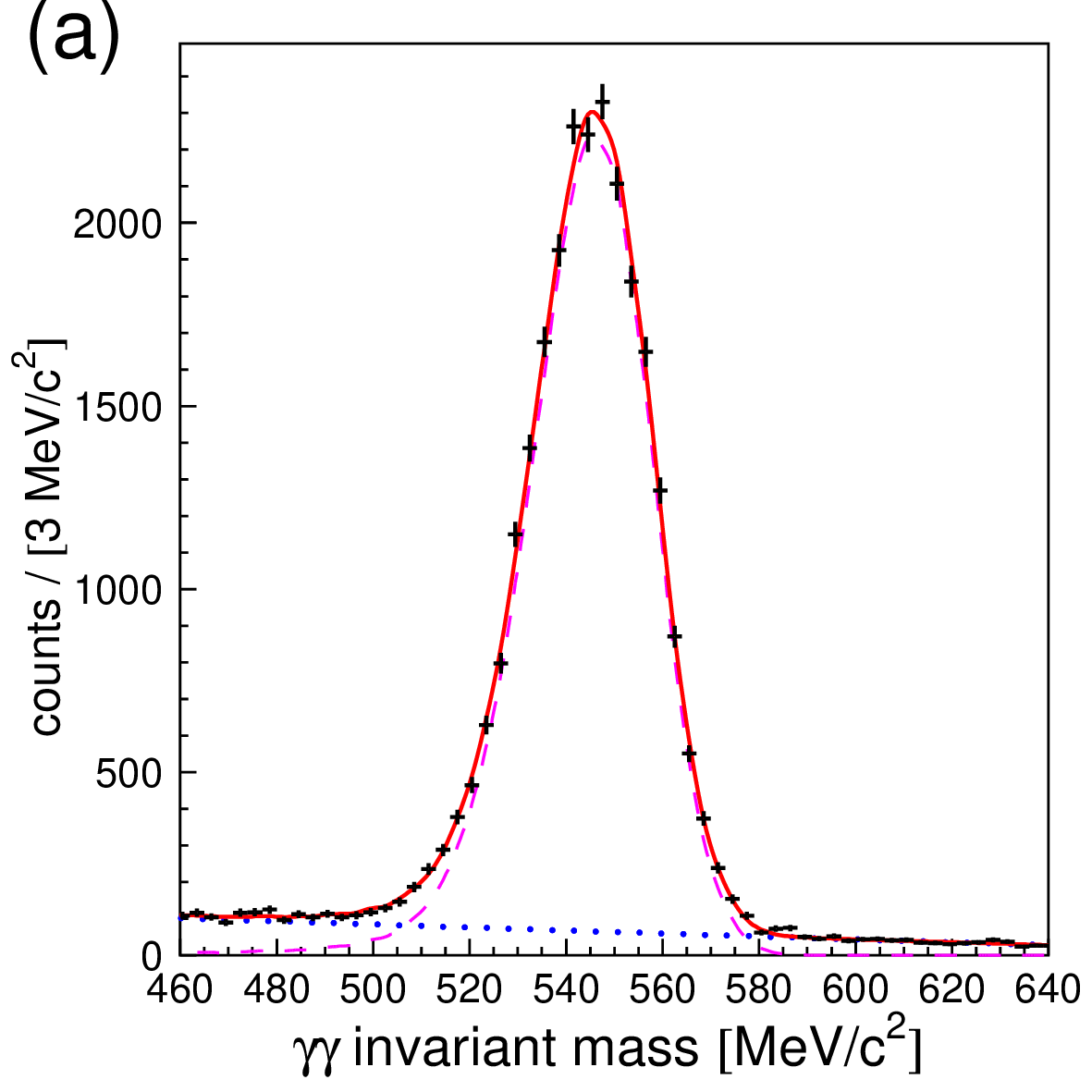}
     \includegraphics[width=4.2cm]{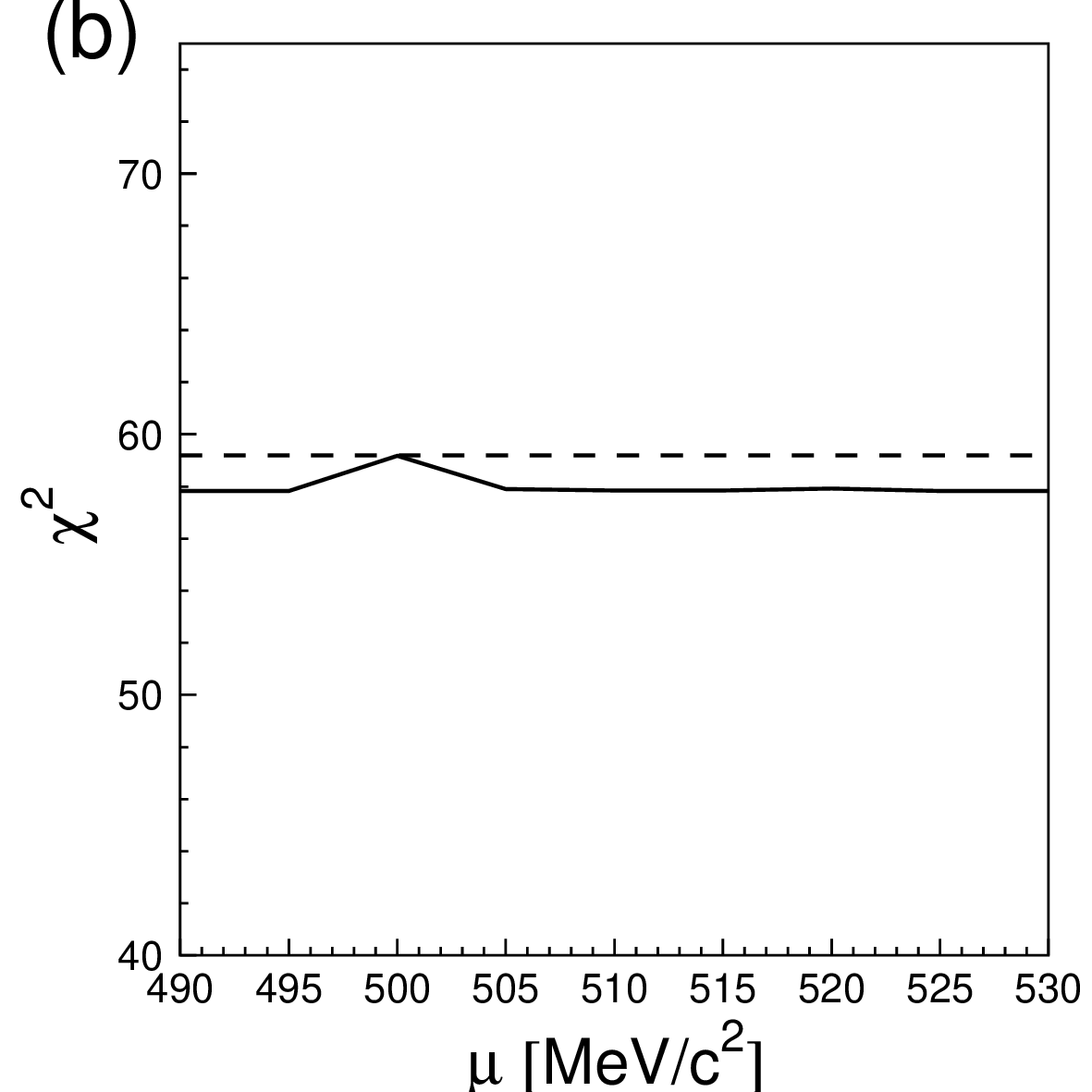}
     \caption{(a) The $\gamma\gamma$ invariant mass spectrum around the $\eta$ mass peak with the fit result (red-solid line). The magenta-dashed and blue-dotted lines represent the template shape for $\eta\to\gamma\gamma$ decays and the polynomial background function, respectively. (b) The $\mu$ dependence of $\chi^2$'s for the fits with Gaussian signal functions (solid line). The dashed line indicates $\chi^2$ for the fit without the signal function.}
    \label{fig:7}
   \end{minipage}
   \hspace{0.1cm}
   \begin{minipage}[]{0.48\textwidth}
    \includegraphics[width=4.2cm]{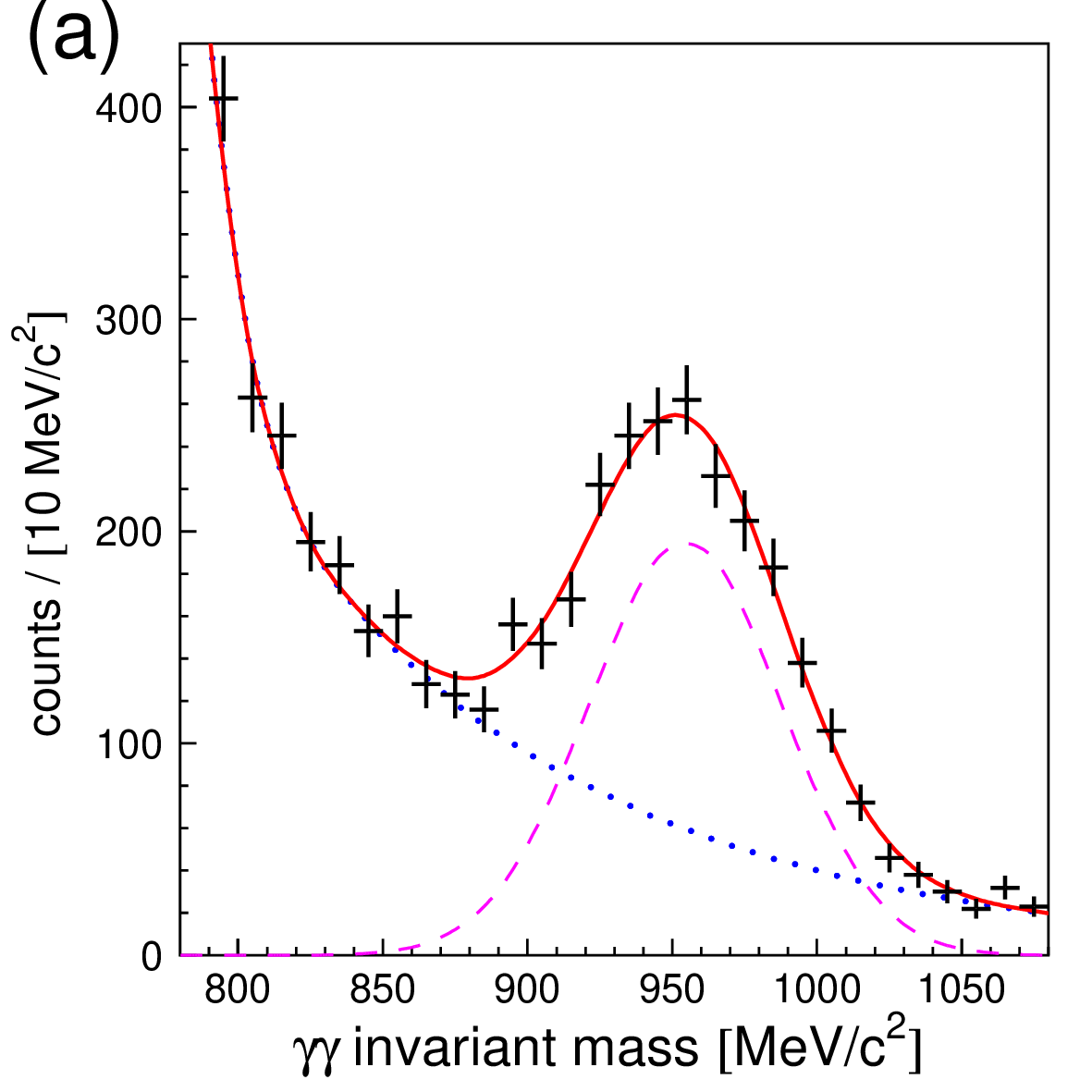}
    \includegraphics[width=4.2cm]{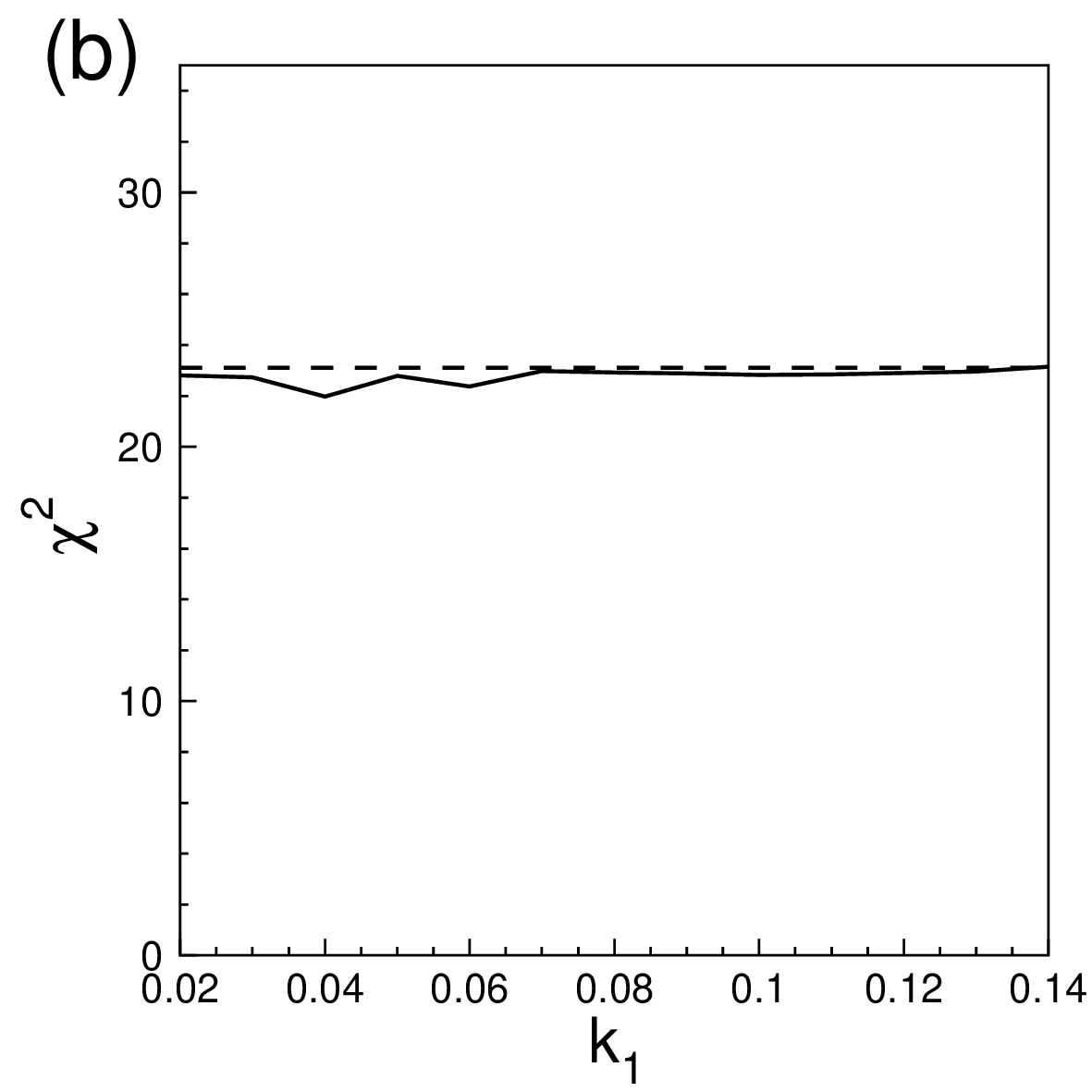}
    \caption{(a) The $\gamma\gamma$ invariant mass spectrum for LH$_2$ target data with the fit result. The notation of lines is the same as that in Fig.~\ref{highpfit} in the main text. (b) The $k_1$ dependence of $\chi^2$'s for the fits with simulated signal functions (solid line). The dashed line indicates $\chi^2$ for the fit without the signal function.}
    \label{fig:8}
   \end{minipage}
  \end{minipage}
 \end{tabular}
\end{figure*}

{\it Spectral fit with a Gaussian signal function.}--
Thirdly, a Gaussian function was assumed to express the $\eta^\prime$ mass reduction signal in the spectral fit.
Figure~\ref{fig:6}(a) shows the best fit result for the low-momentum sample.
In the way similar to the spectral fit with a simulated signal template, the $\chi^2$ value reaches its minimum at the Gaussian mean $\mu$ of $925$~$\mathrm{MeV/c^2}$, providing a significance of $4.0 \sigma$.
The result of $\chi^2$ difference tests is shown as a function of $\mu$ in Fig.~\ref{fig:6}(b) with red-thick lines.
The uncertainty for the best fit value, $\mu = 925$~$\mathrm{MeV/c^2}$, was evaluated to be $-7.9$ and $+6.3$~$\mathrm{MeV/c^2}$ based on the $\mu$ values at $\Delta \chi^2 = 19.4-1$.
The fitting result may exhibit a less mass-reduction of $\eta^\prime$ than that at the normal nuclear density since the signal component includes $\eta^\prime$ decay events at a lower density.
The result of similar tests for the high-momentum sample is also shown with blue-thin lines in Fig.~\ref{fig:6}(b).

{\it Reference analysis for the $\eta$ mass peak.}--
The validity of the realistic MC simulation to reproduce a quasi-free mass peak with $\gamma \gamma$ decays in vacuum was tested using low-momentum $\eta\to\gamma\gamma$ events.
Ten independent samples with $P_{\gamma\gamma}<0.6$~$\mathrm{GeV/c}$, each selected to have statistics comparable to the low-momentum $\etap$ sample, were examined.
Figure~\ref{fig:7}(a) shows a fitting result of the template shape generated by the simulation of $\eta$ photoproduction and the third-order polynomial function to the experimental $\gamma \gamma$ invariant mass distribution.
This fit provides $\chi^2 = 59.2$ for $ndf = 55$, indicating a good consistency.
Figure~\ref{fig:7}(b) shows the $\chi^2$'s for the fits to the sample of Fig.~\ref{fig:7}(a) with the variation of $\mu$ in the case of considering a Gaussian signal component.
For all the examined samples, it was confirmed that the $\chi^2$ difference tests gave signal significances of less than $0.9\sigma$ over the scanned values of $\mu$.
Note that Fig.~\ref{fig:7} shows the case yielding the highest average significance among the ten samples.

{\it Reference analysis using the liquid hydrogen data.}--
The same analysis procedure as the low-momentum sample of the carbon target data was applied to the liquid hydrogen (LH$_2$) target data to confirm no existence of spectral change.
Figure~\ref{fig:8}(a) shows the $\gamma \gamma$ invariant mass distribution for the low-momentum $\eta^\prime$ sample of the liquid hydrogen data.
The red-solid line indicates the fit result using the template shape of quasi-free $\eta^\prime$ peak with $\gamma \gamma$ decays in vacuum and the background functions.
Figure~\ref{fig:8}(b) is the result of $\chi^2$ difference tests using the fit model with simulated signal templates.
No obvious signal of spectral change was seen as expected.

\end{document}